\documentclass[a4paper]{jpconf}
\usepackage{graphicx}
\newtheorem{theorem}{Theorem}
\usepackage{color}
\begin{document}
\title{From Quantum Mechanics to Quantum Field Theory: The Hopf route}
\author{A. I. Solomon{$^1$ $ ^2$}, G. E. H. Duchamp$^3$, P. Blasiak$^4$, A. Horzela$^4$ and K.A. Penson$^2$}
\address{
$^1$ Physics and Astronomy Department,The Open University,
Milton Keynes MK7~6AA, UK}
\address{
$^2$ Lab.de Phys.Th\'eor. de la Mati\`ere Condens\'ee,
University of Paris VI, France}
\address{
$^3$ Institut Galil´ee, LIPN, CNRS UMR 7030 99 Av. J.-B. Clement, F-93430 Villetaneuse,
France}%
\address{
$^4$ H. Niewodnicza{\'n}ski Institute of Nuclear Physics,
Polish Academy of Sciences,
Division of Theoretical Physics,
ul. Eliasza-Radzikowskiego 152, PL 31-342 Krak{\'o}w, Poland}

\ead{a.i.solomon@open.ac.uk, gduchamp2@free.fr, pawel.blasiak@ifj.edu.pl, andrzej.horzela@ifj.edu.pl, penson@lptl.jussieu.fr}

\begin{abstract}
We show that the combinatorial numbers known as {\em Bell numbers} are generic in quantum physics.  This is because they arise in the procedure known as {\em Normal ordering} of bosons, a procedure which is involved in the evaluation of quantum functions such as the canonical partition function of quantum statistical physics, {\it inter alia}.  In fact, we shall show that an evaluation of the non-interacting partition function for a single boson system is identical to integrating the {\em exponential generating function} of the Bell numbers, which is a  device for encapsulating a combinatorial sequence in a single function.

We then introduce a remarkable equality, the Dobinski relation, and use it to indicate why renormalisation is necessary in even the simplest of perturbation expansions for a partition function.

Finally we introduce a global algebraic description of this simple model, giving a Hopf algebra, which provides a starting point for extensions to more complex physical systems.
\end{abstract}
\section{Introduction}
The motivation for this work is as follows: Recent analysis of perturbative quantum field theory (pQFT) has shown that at the heart of this theory there lies a Hopf algebra description.  Since pQFT is a subtle and as yet not completely solved area, it is of pedagogical value  to examine simpler systems to illustrate how these algebraic constructs may arise. In this note we examine  the simple theory given by the quantum partition function $Z$ of a non-relativistic boson system, a completely solvable model (at least in the free boson case). On the basis of a combinatorial methodology, we show that certain combinatorial numbers, known as Bell numbers, play a central role in such an analysis. The graphical approach to the description of these combinatorial numbers leads to a sort of non-relativistic analogue of Feynman diagrams.  En route, we also remark how that even in this simple case, uncritical expansion methods may lead to divergences, which can be cured by regularization; this imitates the renormalization methods of pQFT.  This theory does not depend on space or time; so it is a non-field theory.  However, the description of the Quantum Partition Function that we employ uses the coherent state parameter $z$ as integration variable; and this plays the role of integration over space and time in the relativistic case.

The core of the note then deals with how an elementary Hopf algebra arises naturally from this approach. Although the resulting Hopf algebra, which we call BELL (since it arises directly from a graphical description of the combinatorial Bell numbers), is rather too uncomplicated, (technically it is   commutative and co-commutative), it  serves as a starting point for an extension to more complicated algebraic systems.
Indeed, in later work we show that extensions and deformations of this algebra can emulate that of pQFT, thus supplying an alternate, algebraic, route from the physical path which takes us from the non-relativistic arena to that of relativistic quantum field theory.

\section{Bell and Stirling Numbers; Exponential Generating Function}
The  {\em Bell number} $B(n)$ simply gives the number of partitions of $\{1,2,\ldots,n\}$. Thus for the set $\{1,2,3\}$, we have
$$
\{1,2,3\},\; \; \;  \{1,2\}\{3\}\; \; \; \{1,3\}\{2\}\; \; \; \{3,2\}\{1\}\; \; \; \{1\}\{2\}\{3\}
$$
so that $B(3)=5$.
 Equivalently, $B(n)$ gives the number of ways in which $n$ distinct objects may be distributed among  $n$ identical containers, some of which may remain empty. The first few values of $B(n)$ are
$$
B(n)=1,\,1,\,2,\,5,\,15,\,52,\,203,\,\ldots,\quad n=0,\,1,\,2,\,\ldots\,.
$$

A compact way of defining a combinatorial sequence is by means of a {\em generating function}.  For the Bell numbers, the {\em exponential generating function} is given by
\begin{equation}\label{egf}
\sum_{n=0}^\infty {B(n)}\,\frac{x^n}{n!}= \exp(e^{x}-1).
\end{equation}
We shall give a proof of Eq.(\ref{egf})  after we introduce a graphical description for the Bell numbers in Section \ref{scgt}.
A slight generalization of the exponential generating function for the Bell numbers is given by that defining the {\em Bell polynomials} $B_n(y)$:
\begin{equation}\label{bpgf1}
 \exp\left(y\bigl(e^{x}-1\bigr)\right)=\sum_{n=0}^\infty B_n(y)\,\frac{x^n}{n!}
\end{equation}
With this notation $B(n)\equiv B_n(1)$.

{\em Stirling numbers} (of the second kind) $S(n,k)$
are related to the Bell numbers.
These are defined as the number of ways of putting $n$ different objects
into $k$ identical containers, leaving none empty. From this
definition we have
$B(n)=\sum_{k=1}^n S(n,k)$.

Bell and Stirling numbers are fundamental in quantum theory. This is because they
arise naturally in the {\em normal ordering problem} of bosonic operators.  By the normal ordering operation ${\mathcal N}f(a,a^{\dagger})$ we mean reorder the boson operators in $f(a,a^{\dagger})$ so that all annihilation operators are on the right. For
canonical boson creation and annihilation operators $\{a^{\dagger},a\}$ satisfying  $[a,a^\dag]=1$, the  Stirling numbers of the
second kind $S(n,k)$ intervene through \cite{k1,k2}
\begin{eqnarray}
{\mathcal N}(a^\dag a)^n=\sum_{k=1}^nS(n,k) (a^\dag)^k a^k\,.
\end{eqnarray}
The corresponding Bell numbers $B(n)=\sum_{k=1}^n S(n,k)$ are
simply the expectation values
\begin{equation}\label{Bell}
 B(n)=\langle z|(a^{\dagger}a)^n|z\rangle_{z=1}
\end{equation}
taken in the coherent state defined by
\begin{equation}\label{cs}
 a|z\rangle=z|z\rangle
\end{equation}
for $z=1$.
\section{Graphs}\label{graph}
\subsection{Bell number diagrams}
We now give a graphical representation of the Bell numbers, based on work of Brody, Bender and Meister \cite{ben1,bbm1}, which we have extended in \cite{bbs}.
Consider labelled lines which emanate from a white dot, the
origin, and finish  on a black dot, the vertex. We shall allow
only one line from each white dot but impose no limit on the
number of lines ending on a black dot. Clearly this simulates the
definition of $S(n,k)$ and $B(n)$, with the white dots playing the
role of the distinguishable objects, whence the lines are
labelled, and the black dots that of the indistinguishable
containers. The identification of the graphs for 1, 2 and 3 lines
is given  in Figure 1 below.
\begin{figure}[h]
\hspace{2cm}
\resizebox{10cm}{!}{\includegraphics{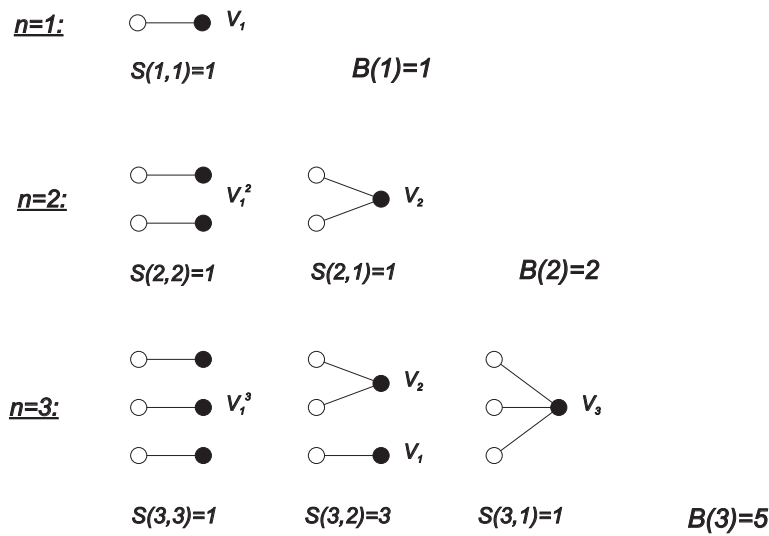}}
\caption{\label{inter}Graphs for $B(n)$, $n=1$, 2, 3.}
\end{figure}
We have concentrated on the Bell number sequence and its
associated graphs since, as we shall show, there is a sense in
which this sequence of graphs is {\em generic} in the evaluation of the Quantum Partition Function. By this we mean the following: The exponential generating function for the  Bell polynomials is identical to the  (integrand of) the Canonical Quantum Partition Function of a {\em non-interacting} single-mode boson model. Then,  when interactions are present these may be incorporated by the use of suitable strengths associated with the vertices of the graphs. That is, we can represent the  perturbation expansion of an interacting model by the same sequence of
graphs as in  Figure 1, with suitable vertex multipliers
(denoted by the $V$ terms in the same figure).
\subsection{The Connected Graph theorem}\label{scgt}
We introduce this important theorem\cite{cgt} in order to obtain the exponential generating function of the Bell numbers.  The theorem states:
\begin{theorem}
Connected Graph Theorem: If ${\cal C}(x)=\sum_{n=1}^{\infty}{c(n)x^n}/{n!}$ is the exponential generating function of a set of CONNECTED labelled graphs then ${\cal A}(x)=\exp{\cal C}(x)$
is the exponential generating function of ALL the graphs.
\end{theorem}

\begin{figure}[h]
\hspace{5cm}
\resizebox{ 5 cm}{!}{\includegraphics{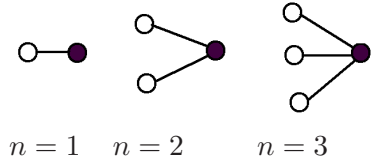}}

\hspace{50 mm} $n=1\;\;\;\;n=2\;\;\;\; \; \; \; \; \; n=3$
\caption{\label{cg}Connected Graphs for $B(n)$, $n=1$, 2, 3.}
\end{figure}

In Figure \ref{cg} we see that there is only {\it one} connected graph for each $n$, so
$${\cal C}(x)= x/1!+x^2/2!+x^3/3!+\ldots = e^x-1$$
whence
$${\cal A}(x)=\exp(e^x-1)$$
gives the exponential generating function for the Bell numbers.

\section{Quantum Partition Function}
We now show that the Canonical  Partition Function of Quantum Statistical mechanics is simply related to the exponential generating function of the Bell polynomials.
\subsection{Free boson gas and Bell polynomials}\label{sub;ho}
The  canonical partition function $Z$ associated with a quantum  mechanical  hamiltonian $H$ is given by
\begin{equation}\label{pf}
Z={\rm Tr}\exp(-\beta H)\,
\end{equation}
where $\beta$ is the usual inverse temperature ( $1/{k_{Boltzmann}T}$).
We start with  the elementary  case of  the hamiltonian for the single--mode free
boson gas $H=\epsilon a^{\dagger}a$ (ignoring an additive
constant), $\epsilon>0$. The usual computation of the partition
function, exploiting the completeness property
$\sum_{n=0}^{\infty}|n\rangle\langle n|=I$, is immediate:
\begin{eqnarray}
Z&=& {\rm Tr}\exp(-\beta \epsilon a^{\dagger}a) \nonumber \\
 &=& \sum_{n=0}^\infty{\langle n|e^{-\beta \epsilon \hat{n}}|n\rangle}  \nonumber \\
 &=& \sum _{n=0}^\infty{e^{-\beta \epsilon n}}   \nonumber \\
 &=&\left(1-e^{-\beta \epsilon}\right)^{-1}\,\label{pf1}.
\end{eqnarray}
However, we may use
{\em any} complete set to perform the trace. We choose coherent
states as defined in Eq.(\ref{cs}) above, which are explicitly given by
\begin{equation}\label{cs2}
|z\rangle= e^{-|z|^2/2}\sum_n ({{z^n}{/n!}) {a^{\dagger}}^n}|0\rangle.
\end{equation}
For these states the completeness or {\em resolution of unity}
property is
\begin{equation}\label{cs1}
 \frac{1}{\pi}\int d^{2}z |z\rangle\langle z|=I\equiv
 \int d\mu(z)|z\rangle\langle z|.
\end{equation}
The appropriate trace calculation is
\begin{eqnarray}\label{tr1}
Z&=&\frac{1}{\pi}\int d^{2}z \langle z|\exp\bigl(-\beta \epsilon a^{\dagger}a\bigr)|z\rangle=\\
&=&\frac{1}{\pi}\int d^{2}z \langle
z|:\exp\bigl(a^{\dagger}a(e^{-\beta\epsilon}-1)\bigr):|z\rangle\,,
\end{eqnarray}
where we have used the following well-known relation
\cite{ks,lou} for the {\em forgetful} normal ordering operator
$:\!f(a, a^{\dagger})\!:$ which means {\it normally order the
creation and annihilation operators in $f$ {\em forgetting}\footnote{This procedure generally  alters the value of the operator to which it is
applied.} the
commutation relation $[a,a^{\dagger}]=1$}:
\begin{equation}\label{no1}
{\mathcal N}exp(xa^{\dagger}a)=:\exp\bigl(a^{\dagger}a(e^{x}-1)\bigr):.
\end{equation}
We therefore obtain, integrating over the angle variable $\theta$
and the radial variable $r=|z|$,
\begin{equation}
Z=\frac{1}{\pi}\int_0^{2\pi}d\theta\int_0^\infty r\,d r
\exp\left(r^2\bigl(e^{-\beta \epsilon}-1\bigr)\right)\,,
\label{z1}
\end{equation}
which gives us $Z=\bigl(1-e^{-\beta\epsilon}\bigr)^{-1}$ as
before.

We rewrite the above equation  to show the connection with our
previously--defined combinatorial numbers. Writing $y=r^2$ and
$x=-\beta \epsilon$, Eq.(\ref{z1}) becomes
\begin{equation}\label{z2}
Z=\int\limits_0^\infty d y\exp\bigl(y(e^{x}-1)\bigr)\,.
\end{equation}
This is an integral over  the classical exponential generating function for
the Bell polynomials as given in Eq.(\ref{bpgf1}). This leads
to the combinatorial form for the partition function
\begin{equation}\label{z3}
Z=\int_0^\infty dy\sum_{n=0}^\infty B_n(y)\,\frac{x^n}{n!}\,.
\end{equation}
\subsection{General partition functions}
We now apply this graphical approach to the general partition
function in second quantized form, with the usual definition for
the partition function Eq.(\ref{pf}).
In general the hamiltonian is given by  $H=\epsilon
w(a,a^{\dagger})$, where $\epsilon$ is the energy scale, and $w$
is a string (= sum of products of positive powers) of boson
creation and annihilation operators. The partition function
integrand $F$ for which we seek to give a graphical expansion, is
\begin{equation}\label{pfz}
 Z(x)=\int{F(x,z)\,d\mu(z)}\,,
\end{equation}
where
\begin{eqnarray}
F(x,z)&=&\langle z|\exp(xw)|z\rangle= \hskip20mm(x=-\beta \epsilon)\nonumber \\
&=&\sum_{n=0}^{\infty}\langle z|w^n|z\rangle\,\frac{x^n}{n!}\nonumber \\
&=&\sum_{n=0}^{\infty}W_n(z)\,\frac{x^n}{n!}\nonumber \\
&=&\exp\biggl(\;\sum_{n=1}^{\infty}V_n(z)\,\frac{x^n}{n!}\biggr),
\end{eqnarray}
with obvious definitions of $W_n$ and $V_n$. The sequences
$\left\{W_n\right\}$ and $\left\{V_n\right\}$ may each be
recursively obtained from the other \cite{pou}. This relates the
sequence of multipliers $\{V_n\}$ of Figure 1 to the hamiltonian
of Eq.(\ref{pf}). The lower limit $1$ in the $V_n$ summation is a
consequence of the normalization of the coherent state
$|z\rangle$.

Although Eq.(\ref{z3}) is remarkably simple in form, it is often
by no means a straightforward matter to evaluate the analogous
integral for other than the free boson system considered here.
Further, as we show below,  we may not interchange the integration
and the summation, as each individual $y$ integral diverges.
At this point the analogue of this model to perturbative quantum field theory becomes more transparent.  For example, to calculate the partition Function $Z(x)$, we must integrate over the Partition Function Integrand $F(x,z)$.  Here, the integration over the coherent state parameter $z$ plays the role of the space-time integration of Quantum Field Theory. As we have just remarked,  with reference to Eq.(\ref{z3}) for the non-interacting case, and as in pQFT (perturbative quantum field theory), term-by-term integration, corresponding to fallaciously interchanging the summation and integration actions, results in infinities at each term, as we now  verify for the non-interacting case.  Thus even here an elementary form of renormalization is necessary, if we  resort to term-wise integration.

Considerations such as these have led many authors, as noted in reference \cite{kre}, to consider a global, algebraic approach to pQFT; and this we  shall do in Section \ref{hopf}.

But first we show how divergences arise.
\section{The Dobi\'nski Relation and Divergence}
\subsection{Divergence}
The celebrated Dobi\'nski relation for the Bell numbers is given by (see, for example, \cite{wilf}):
\begin{eqnarray}\label{Dobinski}
B(n)=\frac{1}{e}\sum_{k=0}^\infty\frac{k^n}{k!}, \ \ \ \ \ \ n=1,2,\ldots
\end{eqnarray}
and for the Bell polynomials by
\begin{eqnarray}\label{dobpoly}
B_{n}(y)&=&e^{-y}\sum_{k=0}^{\infty}{\frac{k^n}{k!}y^k} .
\end{eqnarray}
These relations have been extended in \cite{hier}.

Using the Dobi\'nski relation Eq.(\ref{dobpoly}) we can rewrite the expression for the partition function for the free single-mode boson Eq.(\ref{z3}) as
\begin{eqnarray}\label{z4}
 Z&=&\int_0^\infty dy\sum_{n=0}^\infty B_n(y)\,\frac{x^n}{n!}\\
  &=&\int_0^\infty \sum_{n=0}^{\infty} \frac{x^n}{n!}\{e^{-y}\sum_{k=0}^{\infty}{\frac{k^n}{k!}}y^k\}dy
\end{eqnarray}

We may {\it attempt} to write this as  a perturbation expansion (in $x \equiv - \beta \epsilon$) by interchanging the integration and summation, thus:
\begin{eqnarray}\label{z5}
Z&=&\sum_{n=0}^{\infty} \frac{x^n}{n!}\sum_{k=0}^{\infty}{\frac{k^n}{k!}}\int_0^\infty \{e^{-y}y^k\}dy\\
 &=&\sum_{n=0}^{\infty} \frac{x^n}{n!}\sum_{k=0}^{\infty}{\frac{k^n}{k!}}\Gamma(k+1) \\
 &=&\sum_{n=0}^{\infty} \frac{x^n}{n!}\sum_{k=0}^{\infty}k^n.
\end{eqnarray}
Therefore the resulting ``perturbation expansion'' is of the form
$$ Z=\sum_0^\infty{b_n x^n}\;\; \; \; \; \; \; \; b_n  = {\frac{1}{n!}}\sum_{k=0}^{\infty}{k^n} = \infty $$
with each term in the expansion being infinite.

Thus even in this simple case of a free boson expansion, a perturbation expansion obtained from the illegal interchange of integration and summation is divergent in every term.
\subsection{Regularization}
In this section we illustrate how the method of regularization may be applied to alleviate the problem of divergences.
Writing the partition function Eq.(\ref{z2}) as
\begin{equation}\label{z6}
   Z=\int_{0}^{\infty}e^{-\alpha y} \; dy \; \; \; \; \; \; \; \; (\alpha=1-e^{-\beta \epsilon})
\end{equation}
immediate evaluation gives $Z=1/\alpha$ as in Eq.(\ref{pf1}).

However, if we expand
\begin{equation}
 Z=\int_{0}^{\infty}\sum_{n=0}^{\infty} {{(-\alpha y)^n}/n!} \; dy
\end{equation}
{\em and} exchange the order of integration, we obtain
\begin{equation}
 {\rm (FALSE)}\; \; \; \; \; Z=\sum_{n=0}^{\infty} {(-\alpha)^n/n!}\int_{0}^{\infty}y^n \; dy
\end{equation}
each term of which diverges.
The {\em regularization} method  is to write
\begin{equation}
  Z=\lim_{M \rightarrow \infty} \int_{0}^{M}\sum_{n=0}^{\infty} {{(-\alpha y)^n}/n!} \; dy
\end{equation}
and, since the integration is over a finite range, we may interchange integration and summation, to obtain
\begin{eqnarray}
Z &=& \lim_{M \rightarrow \infty}\sum_{n=0}^{\infty} {{(-\alpha )^n}/n!} \int_{0}^{M} y^n\; dy   \\
&=& \lim_{M \rightarrow \infty}\sum_{n=0}^{\infty}\{(-\alpha )^n/n!\}\{M^{(n+1)}/(n+1)\} \\
&=& \lim_{M \rightarrow \infty}\sum_{n=0}^{\infty}\{(-\alpha M)^{(n+1)}/(n+1)!\}(-1/\alpha)  \\
 &=& \lim_{M \rightarrow \infty}\{(e^{-M\alpha}-1)(-1/\alpha)\} \\
 &=& 1/\alpha
\end{eqnarray}
as before.

Note that $y$ bounded by a finite $M$ implies the eminently reasonable assumption of a maximum energy for the coherent state as defined in Eq.(\ref{cs}).
\section{Hopf Algebra structure}\label{hopf}
So far we have discussed the analytic and combinatorial structure of our simple model.  Modern approaches to perturbative quantum field theory introduce a global algebraic structure.  We describe the analogue of this {\it Hopf Algebra structure} in the context of  our model, which we refer to as BELL below.
We first introduce a basic Hopf Algebra, which we call POLY, generated by a single parameter $x$.  This is essentially a standard one-variable polynomial algebra on which we impose the Hopf operation of co-product, co-unit and antipode, as a useful pedagogical device for describing their properties. The Hopf structure associated with our model described above is simply a multi-variable extension of {\bf POLY}.
\subsection{The Hopf Algebra {\bf POLY}}
{\bf POLY} consists of polynomials in $x$ (say, over the real  field  $\mathcal R$, for example).The standard algebra structure of addition and associative multiplication is obtained in the usual way, by polynomial addition and multiplication.  The additional Hopf operations are:
\begin{enumerate}
\item The coproduct $\Delta:{\bf POLY}{\longrightarrow} {\bf POLY}\times {\bf POLY}$ is defined by
\begin{eqnarray}
\Delta(e)&=&e\times e  \; \; \; \;({\rm unit}\; \; e) \nonumber \\
\Delta(x)&=&x \times e +e \times x  \; \; \; \; ({\rm generator}\; \; x) \nonumber \\
\Delta(AB)&=&\Delta(A)\Delta(B) \; \; \; {\rm otherwise} \nonumber
\end{eqnarray}
so that $\Delta$ is an algebra homomorphism.
\item The co-unit $\epsilon$ satisfies $\epsilon(e)=1$ otherwise $\epsilon(A)=0$.
\item The antipode ${\mathcal S}:{\bf POLY}{\longrightarrow} {\bf POLY}$ satisfies ${\mathcal S}(e)=e$; on the generator $x$, ${\mathcal S}(x)=-x$. It is an {\em anti-homomorphism}, i.e. ${\mathcal S}(AB)={\mathcal S}(B){\mathcal S}(A)$.
\end{enumerate}
It may be shown that the foregoing structure ${\bf POLY}$ satisfies the axioms of a commutative, co-commutative Hopf algebra.
\subsection{{\bf BELL}}
We now briefly describe the  Hopf algebra   ${\bf BELL}$ which is appropriate for the diagram structure introduced in this note, defined by  the diagrams of Figure 1.
\begin{enumerate}
\item Each distinct diagram is an individual basis element of ${\bf BELL}$; thus the dimension is infinite. (Visualise each diagram in a ``box''.) The sum of two diagrams is simply  the two boxes containing the diagrams. Scalar multiples are formal; for example, they may be  provided by the $V$ coefficients.
\item The identity element $e$ is the empty diagram (an empty box).
\item Multiplication is the juxtaposition of two diagrams within the same ``box''. ${\bf BELL}$ is generated by the {\em connected} diagrams; this is a consequence of the Connected Graph Theorem (as discussed above in \ref{scgt}).  Since we have not here specified an order for the juxtaposition, multiplication is commutative.
\item The coproduct $\Delta:{\bf BELL}{\longrightarrow} {\bf BELL}\times {\bf BELL}$ is defined by
\begin{eqnarray}
\Delta(e)&=&e\times e  \; \; \; \;({\rm unit}\; \; e) \nonumber \\
\Delta(y)&=&y \times e +e \times y  \; \; \; \; ({\rm generator}\; \; y) \nonumber \\
\Delta(AB)&=&\Delta(A)\Delta(B) \; \; \; {\rm otherwise} \nonumber
\end{eqnarray}
so that $\Delta$ is an algebra homomorphism.
\item The co-unit $\epsilon$ satisfies $\epsilon(e)=1$ otherwise $\epsilon(A)=0$.
\item The antipode ${\mathcal S}:{\bf BELL}{\longrightarrow} {\bf BELL}$ satisfies ${\mathcal S}(e)=e$; on a generator $x$, ${\mathcal S}(y)=-y$. It is an {\em anti-homomorphism}, i.e. ${\mathcal S}(AB)={\mathcal S}(B){\mathcal S}(A)$.
\end{enumerate}
\subsection{{\bf BELL} as an extension of {\bf POLY}}
It may be seen that ${\bf BELL}$ is a multivariable version of ${\bf POLY}$. To show this, we  code the diagrams by letters. We use an infinite alphabet $Y=\{y_1,y_2,y_3\cdots \}=\{y_k\}_{k\geq 1}$ and code each connected diagram with one black spot and $k$ white spots with the letter $y_k$. An unconnected diagram will be coded by the product of its letters.\\
In this way, the diagrams of Figure 1 are coded as follows:
\begin{itemize}
	\item first line : $y_1$
	\item second line : $y_1^2$ and $y_2$
	\item third line : $y_1^3$ and $y_1y_2$ and $y_3$\ .
\end{itemize}
Thus one sees that each diagram of weight $n$ with $k$ connected components is coded bijectively by a monomial
of weight $n$ (the weight of a monomial $y_{i_1}y_{i_2}\cdots y_{i_r}$ is just the sum of the indices $\sum_{j=1}^r i_j$) and $k$ letters. The algebra ${\bf BELL}$ is coded by commutative polynomials in the infinite alphabet $Y$; that is, the coding is an isomorphism ${\bf BELL}\rightarrow {\mathcal R}[Y]$. As an aside, one may note that  the basis elements of ${\bf BELL}$ are sometimes referred to as {\em forests}.
As above, it  may be shown that the foregoing structure ${\bf BELL}$ satisfies the axioms of a commutative, co-commutative Hopf algebra.
\section{Discussion}
In this note we have shown the value of the combinatorial method, which leads {\it inter alia} through the graphical approach, to a Hopf algebra description of a simple non-relativistic free boson system. We named the resulting (commutative, co-commutative) Hopf  algebra BELL.  To extend this simple description to more realistic models, it is convenient to introduce a symmetrized form of BELL; i.e. where the black and white spots appear symmetrically\cite{feynman}. Thus we have introduced the Hopf algebra DIAG where a typical series of graphs is given in Figure \ref{diag} below.
\begin{figure}[h]
\hspace{2cm}
\resizebox{10cm}{!}{\includegraphics{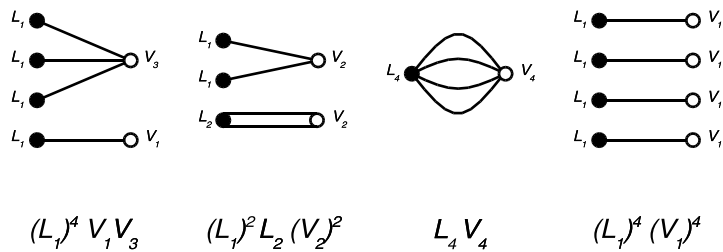}}
\caption{\label{diag}Some Graphs of DIAG}
\end{figure}
Note that since the black and white spots play symmetrical roles, in the figure we permit ourselves to exchange their order from that used in, for example, Figure \ref{inter}.  We may also apply ``weights'' $L_m$ to the black spots as well.  Although the resulting Hopf algebra DIAG is still commutative/co-commutative,  it may in turn be extended to a non-commutative Hopf algebra (LDIAG) which forms the basis of an extension of this algebraic approach to a description sufficiently general to emulate those algebras which describe pQFT. This extension is described in detail in a subsequent artcle in this conference \cite{MDSG}.
\section*{Acknowlegements} The authors wish to acknowledge support from the Agence Nationale de la Recherche (Paris, France) under Program No. ANR-08-BLAN-0243-2 and from PAN/CNRS Project PICS No.4339(2008-2010) as well as the Polish Ministry of Science and Higher Education Grant No.202 10732/2832.

\end{document}